 \documentstyle[12pt]{article}
\begin{document}

\begin{center}
{\bf \large THE OBSERVATION OF  MULTI-QUARK STRANGE METASTABLE AND
STABLE STATES}

\vskip 5mm P.Z. Aslanyan$^{1,2}$, V.N. Emelyanenko$^1$,
G.G.Rikhvitzkaya$^1$

 \vskip 5mm {\small (1) {\it Joint Institute for Nuclear Research}}
\\
(2) {\it Yerevan State University }
\\
$\dag$ {\it E-mail: petros@spp.jinr.ru}
\end{center}


\begin{abstract} In the metastable state sector the $\Lambda p$,
$\Lambda\pi\pi$ and $\Lambda p \pi$  resonances found in
n+$^{12}C$- collisions at 7 GeV/c. Preliminary results show that
the predicted peaks  has been confirmed in $p^{12}C$ collision at
10 GeV/c. In the stable state sector candidates for S=-2 light and
heavy  dibaryons observed in p+$^{12}C$ collision at 10 GeV/c.
\end{abstract}

\section{Introduction}
 Quark models with confinement predict the existence of exotic
objects as, for instance, glueballs, hybrid states of quarks and
gluons as well as multiquark hadrons \cite{1}- \cite{6}. It
is important to stress that the existence of
 multiquarks is a necessary consequence of modern ideas of non-perturbative
 QCD vacuum and is a matter of principle to comprehend the characteristics of
matter both at microscopic and cosmological levels. New particles
or states of matter containing 2-or more strange quarks have
inspired a lot of experiments at  BNL(AGS), CERN, FNAL, SEBAF, KEK
and JINR.

Stable and metastable strange  dibaryons were searched a long time
ago at LHE JINR, too. A thorough analysis of the merits and
demerits of the available detection techniques and data analysis
procedures has convinced us that the propane  bubble chamber
technique is most adequate to the physics problem at the present
reconnaissance stage. A reliable identification of dibaryons needs
a multivertex kinematic analysis which is in turn is feasible only
using 4$\pi$-detectors and high precision measurements of the
sought objects.

The investigations of multiquark states have been carried out in
two directions.

\subsection{}
 Multiquark resonance formation via the compression mechanism
reduces to the phase transition  of normal density super strange
hadronic matter revealing itself as multiquark resonances [1-3].

The problem of experimental examination of multibaryon strange
resonances was started at LHE from 1962 up to now. The effective
mass spectra of 17 strange multiquark systems were studied for
neutron exposure n $^{12}C\to \Lambda$X at average momentum
$<p_n>$=7.0 GeV/c, and our group succeeded in finding
resonance-like peaks [7-10](Table~\ref{res}) only in five of them
$\Lambda p$, $\Lambda p \pi$, $\Lambda \Lambda$, $\Lambda \Lambda
p $, $\Lambda \pi^+ \pi^+$. Most significant evidence is included
in the Review of Particle Properties.

Up to now, the same group is going on to collect statistics(more 3
time) for strange $V^0$ particles on the photographs of the JINR
2m propane bubble chamber exposed to a 10 GeV/c proton beam because
it is the necessary to improve a statistical significance of the
identified resonance peaks  and to search  for new  strange
multiquark resonance states.  Fig.1(a,b,c) show the preliminary
experimental effective-mass spectrum of the $\Lambda \pi$,
$\Lambda p$ and $\Lambda p \pi$  produced by pC interaction.

\subsection{}
 A simple consideration of symmetry and the properties of
color-magnetic interactions  argues in flavor of increasing
binding in three flavor matter-systems containing u,d and s
quarks. These objects known an H particles(uuddss). This was first
proposed by Jaffe in 1977 using the MIT bag model [3-6]. The
possibility that H dibaryon matter may exist in the core of a
neutron star was also pointed out. The estimated binding energy of
the H particle is  model - dependent and ranges from positive
(unbound) to negative strong bound states( -650MeV). The heavy
isotriplet stable dibaryon H (I=1, J=0+,Y=0,B=2, S=-2) of a 2370
$MeV/c^2$ mass is predicted by the soliton Skirme-like model
[1,3].

The search for weak decay channels for stable S=-2,3 dibaryon
states is being continued to date. A few events, detected on the
photographs of the propane bubble chamber exposed to a 10 GeV/c
proton beam, were interpreted as H dibaryons [11-16]. There are
two groups of events interpreted as S=-2 stable dibaryons(
Table~\ref{eff} ):\\ a) The first group  is formed of three
neutral, S=-2 stable dibaryons, the masses of which are below
$\Lambda\Lambda$ threshold;\\ b) The second group  is formed of
neutral and positively charged S=-2 heavy stable dibaryons \\
(Figs.2,3). The masses of all the three dibaryons coincide within
the errors are over the $\Lambda\Lambda$, $\Xi N$, $\Lambda
\Sigma$ threshold.

\begin{table}
\caption{The effective mass spectra in collisions of 4.0 GeV/c
$\pi^-$ and neutrons of a 7.0 GeV/c average momentum with $^{12}C$
nuclei, have led to the discovery of the peaks presented below.}
\label{res}
\begin{tabular}{|c|c|c|c|c|c|c|c|}  \hline
Resonanse &M&$\Gamma$&Signifi-&$\sigma$&
\multicolumn{2}{|c|}{\underline{Bag model predictions}} \\
system&MeV/$c^2$&MeV/$c^2$&cance&($\mu b/^{12}C$)&M& $J^p$\\
 &&&$N_{sd}$&&(MeV/$c^2$)&\\ \hline
$\Lambda
p$&2095.0$\pm$2.0&7.0$\pm$2.0&5.70$\pm$1.20&55.0$\pm$16.0&2110&$1^-$
\\ $\Lambda
p$&2181.0$\pm$2.0&3.2$\pm$0.5&4.36$\pm$1.21&60.0$\pm$15.0&2169&$1^+$
\\ $\Lambda
p$&2223.6$\pm$1.8&22.0$\pm$1.9&6.24$\pm$1.23&40.0$\pm$12.0&2230&$0^+$
\\ $\Lambda
p$&2263.0$\pm$3.0&15.6$\pm$2.3&8.55$\pm$1.35&85.3$\pm$20.0&2241&$2^+$
\\ $\Lambda
p$&2356.0$\pm$4.0&98.6$\pm$2.5&13.81$\pm$1.39&65.0$\pm$17.0&2253&$1^+$
\\ $\Lambda
p$&2129.2$\pm$0.3&0.7$\pm$0.16&11.37$\pm$1.37&90.0$\pm$20.0&2128&$1^+$\\\hline
$\Lambda
p\pi^{\pm}$&2495.2$\pm$8.7&204.5$\pm$5.6&12.86$\pm$1.68&70.5$\pm$15.0&2500&$0^-,1^-,2^-$
\\ \hline
$\Lambda\Lambda$&2365.3$\pm$9.6&47.2$\pm$15.1&4.2$\pm$1.40&24.2$\pm$7.0&2365&$-.-$
\\ \hline $\Lambda\Lambda
p$&$\approx$3568.3&$<$60&-.-&16.1$\pm$5.2&3570&$5/2^+$ \\  \hline
$\Lambda
\pi^+\pi^+$&1704.9$\pm$0.9&18.0$\pm$0.5&5.3$\pm$1.6&19.0$\pm$0.6&1710&$1/2^-$
\\ $\Lambda
\pi^+\pi^+$&2071.6$\pm$4.0&172.9$\pm$12.4&10.3$\pm$1.5&88.0$\pm$27.0&2120&$1/2^-$
\\ $\Lambda \pi^+\pi^+$&2604.8$\pm$4.8&
85.9$\pm$21.5&5.2$\pm$1.4&31.9$\pm$9.0&2615&$3/2^-$ \\ \hline
\end{tabular}
\end{table}

\begin{table}
\caption{Mass and weak decay channels for the registration of
dibaryons.} \label{eff}
\begin{tabular}{|c|c|c|c|c|c|c|c|}  \hline
N& Channel of decay& Mass H&C.L.&References \\
 &&$(MeV/c^2)$&of fit& \\
 &&Dibaryon&$\%$ &\\ \hline
1&$H^0 \to \Sigma^-p$&$2172\pm15$&$ 99$&Z.Phys.C  39, 151(1988).
\\   \hline 2&$H^0 \to \Sigma^-p,\Sigma^-\to n\pi^-$&$2146\pm1$&
30& JINR RC,\\ &$H^0_1 \to H^0(2146)\gamma$&$2203\pm6$& 51&N
1(69)-95-61,1995. \\   \hline 3&$H^0 \to \Sigma^-p,\Sigma^-\to
n\pi^-$&$2218\pm12$& 69&Phys.Lett B235(1990),208. \\   \hline
4&$H^0 \to \Sigma^-p,\Sigma^-\to n\pi^-$&$2385\pm31$&34&Phys.Lett
B316(1993),593.\\     \hline 5&$H^+ \to p\pi^0\Lambda
^0,\Lambda^0\to p\pi^-$&$2376\pm10$&87&Phys.Lett B316(1993),593
\\\hline 6&$H^+ \to p\pi^0\Lambda ^0,\Lambda^0\to p\pi^-$
&$2580\pm108$&86&Nucl.Phys.75B(1999),63. \\
 &$H^+n \to \Sigma^ +\Lambda ^0~n, \Lambda^0\to p\pi^-$ &$2410\pm90$&6&\\ \hline
7&$H^+ \to p\gamma\Lambda ^0,\Lambda^0\to p\pi^-$
&$2448\pm47$&73&JINR Com.(2002)\\ &$H^+ \to p\pi^0\Lambda
^0,\Lambda^0\to p\pi^-$ &$2488\pm48$&72&E1-2001-265 \\ \hline
\end{tabular}
\end{table}

\newpage

\end{document}